\documentclass[aps,pra,twocolumn,groupedaddress, longbibliography]{revtex4-1}

\usepackage{graphicx}
\usepackage{stmaryrd}
\usepackage{color}
\usepackage{hyperref}

\definecolor{blue_n}{rgb}{0.,0.3,0.5}

\hypersetup{
backref=true, 
pagebackref=true,
hyperindex=true, 
colorlinks=true, 
breaklinks=true, 
citecolor = blue_n, 
urlcolor= blue_n,
linkcolor= black, 
pdftitle={Quantum frequency conversion to telecom of single photons from a nitrogen-vacancy center in diamond}, 
pdfauthor={Ana\"is Dr\'eau, Anna Tchebotareva, Aboubakr El Mahdaoui, Cristian Bonato and Ronald Hanson } 
}

\begin{document}

\title{
Quantum frequency conversion to telecom of single photons\\ from a nitrogen-vacancy center in diamond}

\author{Ana\"is Dr\'eau$^{1,2,3}$}
\email[]{anais.dreau@umontpellier.fr}
\author{Anna Tchebotareva$^{1,4}$}
\author{Aboubakr El Mahdaoui$^{1,2}$}
\author{Cristian Bonato$^{1,2}$}
\email[current address: Institute of Photonics and Quantum Sciences, SUPA, Heriot-Watt University, Edinburgh EH14 4AS, UK]{}
\author{Ronald Hanson$^{1,2}$}

\affiliation{$^1$QuTech, Delft University of Technology, P.O. Box 5046, 2600 GA Delft, The Netherlands}
\affiliation{$^2$Kavli Institute of Nanoscience, Delft University of Technology, P.O. Box 5046, 2600 GA Delft, The Netherlands}
\affiliation{$^3$Laboratoire Charles Coulomb, Universit\'e de Montpellier and CNRS, 34095 Montpellier, France}
\affiliation{$^4$Netherlands Organisation for Applied Scientific Research (TNO), P.O. Box 155, 2600 AD Delft, The Netherlands
}

\begin{abstract}
We report on the conversion to telecom wavelength of single photons emitted by a nitrogen-vacancy (NV) defect in diamond.
By means of difference frequency generation, we convert spin-selective photons at 637 nm, associated with the coherent NV zero-phonon-line, to the target wavelength of 1588 nm in the L-telecom band. The successful conversion is evidenced by time-resolved detection revealing a telecom photon lifetime identical to that of the original 637 nm photon. Furthermore, we show by second-order correlation measurements that the single-photon statistics are preserved. The overall efficiency of this one-step conversion reaches 17\% in our current setup, along with a signal-to-noise ratio of $\approx$7 despite the low probability $(< 10^{-3})$ of an incident 637 nm photon. This result shows the potential for efficient telecom photon - NV center interfaces and marks an important step towards future long-range entanglement-based quantum networks.

\end{abstract}

\maketitle

The realization of a quantum network capable of distributing and processing entanglement across distant nodes is a major goal in quantum technology.
The architecture of such a quantum network relies on connecting many optically active qubit-hosting nodes, separated by tens or hundreds of kilometres, through exchange of quantum information carried by photons propagating in optical fibers \cite{kimble_quantum_2008}. 
Several qubit platforms, such as trapped ions \cite{hucul_modular_2015}, trapped atoms~\cite{ritter_elementary_2012,hofmann_heralded_2012} and quantum dots \cite{gao_coherent_2015, delteil_generation_2016,stockill_phase-tuned_2017} have started demonstrating the building blocks of such a network.

One of the most promising platforms for quantum network nodes is the nitrogen-vacancy (NV) center in diamond. The NV center features a well-controlled electron spin qubit ($S = 1$) that exhibits second-long coherence time \cite{bar-gill_solid-state_2013, abobeih_one-second_2018}.
At cryogenic temperatures, the spin-selective optical transitions can be individually addressed, which has enabled single-shot spin readout~\cite{robledo_high-fidelity_2011} and the generation of spin-photon~\cite{togan_quantum_2010} and spin-spin entanglement ~\cite{bernien_heralded_2013}.
The NV qubit is surrounded by long-lived $^{13}$C nuclear spins \cite{smeltzer_13_2011,dreau_high-resolution_2012} that can be used as additional network qubits, for instance for quantum error correction~\cite{waldherr_quantum_2014, taminiau_universal_2014, cramer_repeated_2016} or remote entanglement distillation~\cite{kalb_entanglement_2017}. The current goal is to combine all those elementary building blocks to demonstrate the practical implementation of an entanglement-based quantum network supported by NV centers in diamond~\cite{dam_multiplexed_2017}. 

However, the optical emission properties of NV centers form a major hurdle towards large-distance quantum networks. In particular, the NV center emits its coherent photons -  those that are useful for generating remote entanglement -  at 637~nm, a wavelength that undergoes high propagation loss in optical fibers ($\approx$ 8 dB/km). For instance, in the recent demonstration of kilometer-scale entanglement between two NV centers in diamond~ \cite{hensen_loophole-free_2015}, this fiber attenuation reduced the entanglement generation rate by a factor of 30 in comparison to local implementations.

\begin{figure*}[t!]
\begin{centering}
\includegraphics[width=1.7\columnwidth]{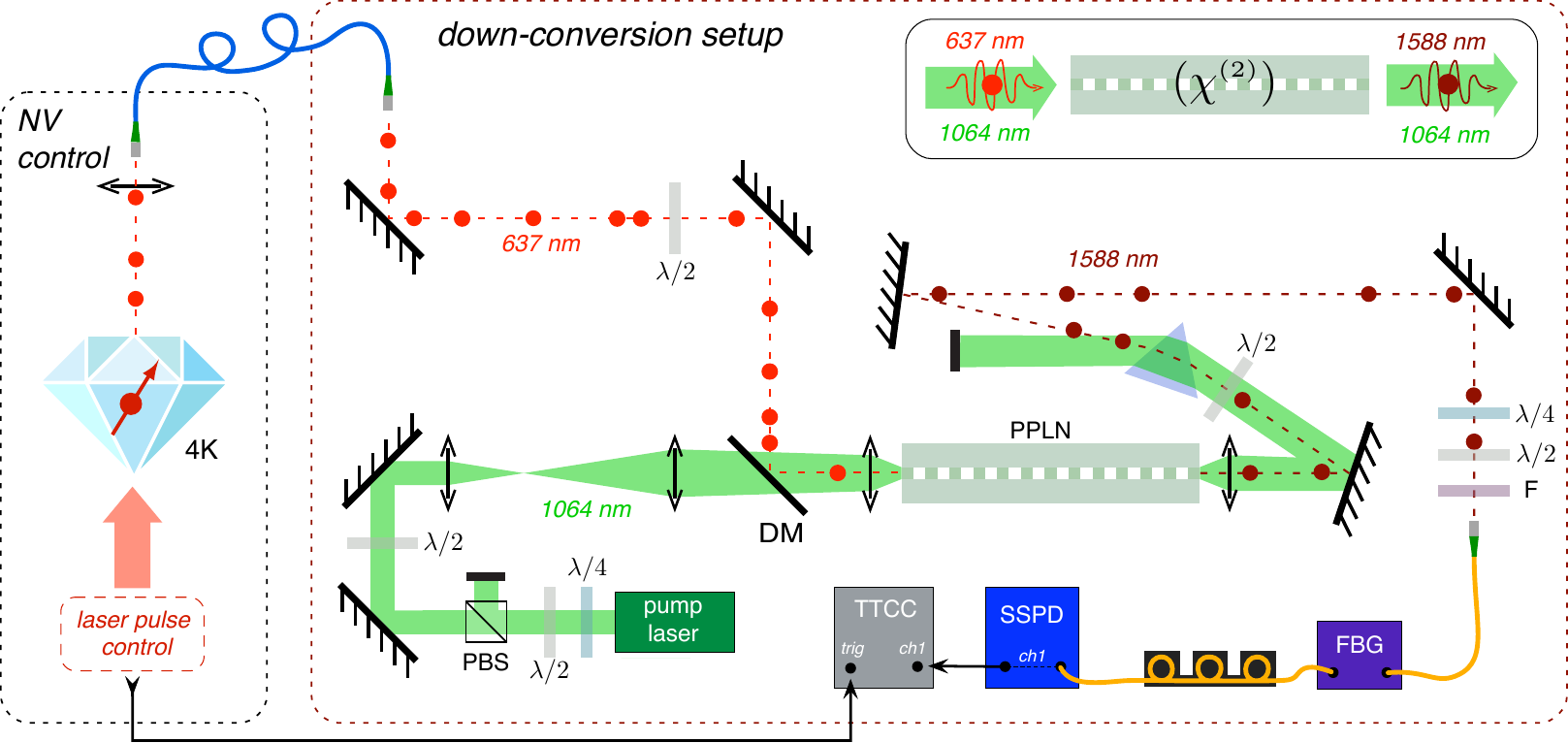}
\caption{
Experimental setup for the conversion of single NV center photons to a telecom frequency. 
Inset depicts the working principle: using a strong pump beam at 1064~nm, an NV center photon emitted at 637~nm is down-converted by difference frequency generation to a 1588~nm photon.
DM: dichroic mirror; PBS: polarizing-beamsplitter; PPLN: periodically-poled lithium niobate non-linear crystal; F: filter ; FBG: fiber-Bragg grating; SSPD: superconducting single photon detector; TTCC: time-tagged correlation card. 
Polarization maintaining and single mode fibers are respectively displayed in blue and yellow. }
\label{fig:setup}
\end{centering}
\end{figure*}

The key to removing this hurdle is to operate the spin-photon interface in the telecom wavelength range, where optical losses in fibers are minimal ($< 0.2$~dB/km \cite{miya_ultimate_1979, nagayama_ultra-low-loss_2002}). 
A promising approach is to down-convert the spin-entangled photons emitted by the NV center using the nonlinear optical process of difference frequency generation (DFG). 
Nonlinear frequency conversion processes applied on quantum light, commonly described as quantum frequency conversion (QFC), can retain the quantum information including entanglement encoded on the initial light \cite{tanzilli_photonic_2005, de_greve_quantum-dot_2012, ikuta_wide-band_2011,ramelow_polarization-entanglement-conserving_2012}. 
Successful down-conversion of single photons from atomic memories \cite{radnaev_quantum_2010, farrera_nonclassical_2016, maring_photonic_2017}, quantum dots \cite{rakher_quantum_2010,ates_two-photon_2012,de_greve_quantum-dot_2012,zaske_visible--telecom_2012} and trapped ions ~\cite{bock_high-fidelity_2017} have recently been demonstrated.

Applying quantum frequency conversion to the NV center is a highly challenging task. First, only the coherent zero-phonon-line (ZPL) emission (corresponding to $\approx 3\%$ of the total emission) \cite{johnson_tunable_2015, riedel_deterministic_2017} can be used to establish entanglement \cite{bernien_heralded_2013}, leading to a low incident photon rate for conversion. 
Significant noise introduced in the nonlinear conversion process by the pump laser, through spontaneous parametric down-conversion (SPDC)~\cite{pelc_influence_2010} or Stokes Raman scattering \cite{ikuta_frequency_2014}, may thus prevent the signal-to-noise ratio at the output from exceeding unity. Second, the solution of operating the pump at a wavelength sufficiently above the target wavelength \cite{pelc_influence_2010} is impeded by the short wavelength (637~nm) of the NV center ZPL. These difficulties are highlighted by a recent experiment on down-converting attenuated 637~nm laser light ~\cite{ikuta_frequency_2014} to the C and L telecombands, 
which achieved an internal conversion efficiency of 44\% but at the cost of a noise photon rate of $\approx 4$Mcts/s - about 3 orders of magnitude above typical ZPL count rates under off-resonant excitation for a NV center embedded in a diamond solid-immersion lense \cite{pfaff_unconditional_2014}.

Here, we overcome these challenges by employing pulsed resonant excitation of the NV center with time-synchronized photon detection, in combination with efficient down-conversion and noise filtering. Through this, we are able to unambiguously demonstrate the down-conversion of single NV center ZPL photons to telecom wavelength. We perform lifetime measurements on the initial and frequency-converted NV center single photons and we assess the conversion efficiency and the noise levels in our conversion setup. Finally, we perform a Hanbury-Brown and Twiss experiment before and after conversion to demonstrate the preservation of the single-photon light statistics. 

\section{Quantum frequency conversion and noise filtering setup}

\begin{figure*}[t!]
\begin{centering}
\includegraphics[width=1.7\columnwidth]{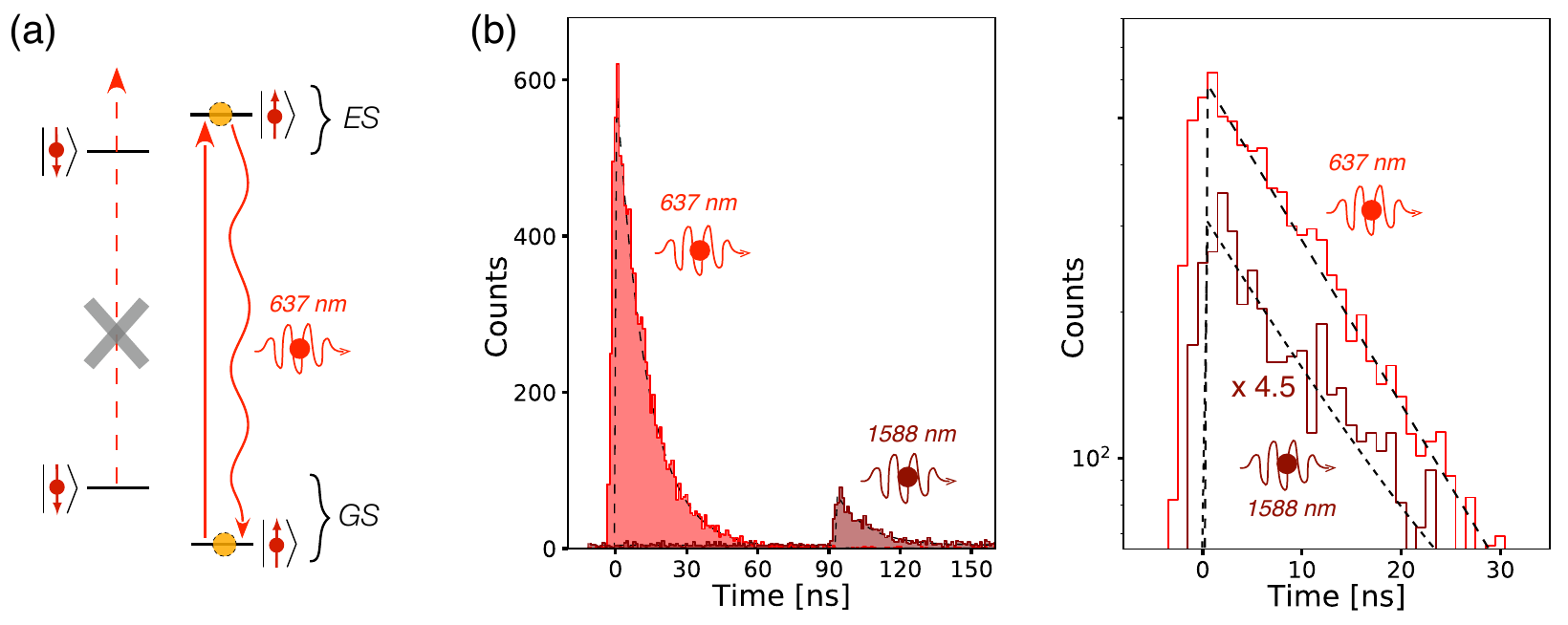}
\caption{
(a) Simplified energy level structure of the relevant NV center optical transitions. 
The NV center is excited resonantly by an optical $\pi$-pulse on a spin-selective optical transition, such that - conditional on the spin state - a single photon is emitted. 
(b) Histograms of the detected counts before and after down-conversion, plotted in linear (left) and logarithmic (right) scales. The sequence was repeated 7.5 million times in both cases. 
The dash lines are fits with an exponential decay function including an offset for the background noise, yielding decay times of $12.9 \pm 1.0$~ns (637~nm) and $12.6 \pm 1.2$~ns (1588~nm).
On the logarithmic plot, the 1588~nm data is shifted in time and its counts multiplied by 4.5 to ease the visual comparison to the 637~nm data.}
\label{fig:tail}
\end{centering}
\end{figure*}

The experimental setup optimized for down-conversion of the NV center single photons is presented in Figure \ref{fig:setup}. 
The photons emitted by a single NV center are extracted from the diamond and coupled into a polarization-maintaining (PM) fiber whose output is directed to the down-conversion stage. 
By means of a Zn-doped type-0 PPLN crystal (NTT Electronics), we use the DFG process to down-convert the NV photons at 637~nm to the telecom wavelength of 1588~nm, using a continuous-wave pump laser at 1064~nm (Coherent Mephisto) (inset of Fig.\ref{fig:setup}). 
The high DFG efficiency required for frequency conversion at the single photon level is achieved by strong optical confinement, for both the NV centre and pump lights, inside a waveguide diced in the PPLN crystal (dimensions: 8$\mu$m x 8$\mu$m x 48mm).
Both input and output facets of the crystal are coated with an anti-reflection layer. 
In- and out-coupling of light from the crystal is realized by using aspherical lenses. 
The coupling efficiency of the NV center photons into the crystal waveguide is estimated to be at least of $\approx 90\%$ by measuring the transmission of a CW laser (Toptica DLPro) at the same wavelength as the NV emission. 
We use a 2-lens telescope in the pump beam path in order to correct both for the chromatic aberration of the crystal in-coupling lens and to achieve a good spatial mode matching between the NV center photon and pump modes inside the waveguide that is not fully single-mode for these wavelengths. 
The quasi-phase matching condition essential for good conversion efficiency is reached by tuning the temperature of the PPLN crystal.

The input pump power sent through the crystal is tuned by adjusting a half-wave plate in front of a polarizing beam splitter. 
At the PPLN crystal output, three steps of spectral filtering are implemented to efficiently extract the down-converted photons from the high background noise coming mostly from the remaining pump photons and the unwanted noise generated during the DFG process. 
First, the telecom photons are split from the pump light by using a dispersive prism. 
Then, they pass through a long-pass filter (Semrock BLP01-1550R-25) before entering a single mode optical fiber.
The last filtering element is a tunable fiber Bragg grating (FBG) bandpass filter (AOS GmbH), featuring a 4~pm ($\approx500$~MHz) bandwidth and a noise rejection of 55~dB. 
The down-converted single photons are finally sent to a superconducting single photon detector (SSPD) (SingleQuantum, jitter $<\!70\,$ps). 
The SSPD quantum efficiency (QE) can be tuned to more than 60\% by changing the bias current applied to the superconducting nanowire, with higher dark-count rates for higher QEs~\cite{natarajan_superconducting_2012}. 
As a compromise, the QE was set to 41~\% for all the data presented here. 
Counts are recorded on a time-tagged correlation card (Picoquant HydraHarp 400, time resolution 1ps), synchronized to the optical excitation of the NV center with a time jitter $<\!80\,$ps.
The optical polarization is controlled at each stage of the setup, to optimize the conversion efficiency, the filtering and the detection efficiency.

\section{Experimental frequency conversion of single NV photons}

We first verify that the photo-detection events at telecom wavelength result from the down-converted photons originating from the NV center in diamond by performing time-resolved lifetime measurements. 
To comply with future quantum networking applications, we use a spin-selective optical transition of the NV center that has previously been used to establish remote entanglement~\cite{bernien_heralded_2013} (Fig.\ref{fig:tail}(a)). After preparation of its optical and spin state ($m_s =0$), the NV center is resonantly driven by a 2-ns optical $\pi$-pulse to the $E_x$ excited state~\cite{bernien_heralded_2013}, after which spontaneous emission of a photon brings the NV back to the ground state.
We first record the NV center ZPL photons before conversion (i.e. at 637~nm) by connecting the input PM fiber directly to a visible avalanche photodiode (LaserComponents COUNT-10C-FC) with specified detection efficiency around $80\%$.
The temporal histogram of counts (Fig.\ref{fig:tail}(b), left panel) exhibits the expected exponential decay with a time constant of $12.9 \pm 1.0$~ns, corresponding to the NV excited state lifetime for the $E_x$ state \cite{goldman_phonon-induced_2015}.

Next, we perform the same time-resolved measurement after frequency conversion by detecting the telecom photons (now at 1588~nm) on the SSPD (Fig.\ref{fig:tail}(b), left panel). Telecom photons are detected at an $\approx$ 90~ns time delay from the excitation pulse, which matches the additional path length for the down-converted photons combined with the electronic delays of the SSPD driver.
As can be seen on the logarithmic plot in Figure \ref{fig:tail}(b), the histogram of counts at telecom wavelength shows the same temporal shape as the initial one for counts at 637 nm, with an exponential time constant of $12.6 \pm 1.2$~ns. 
These results demonstrate that the recorded telecom photons indeed originate from the NV center emission. 
We point out that the histograms of Figure \ref{fig:tail}(b) display the recorded number of detection events; these histograms are therefore not corrected for their respective detector's quantum efficiency.

\section{Quantum frequency conversion efficiency and noise}
\label{sec:eff_noise}

\begin{figure}[h!]
\begin{center}
\includegraphics[width=8cm]{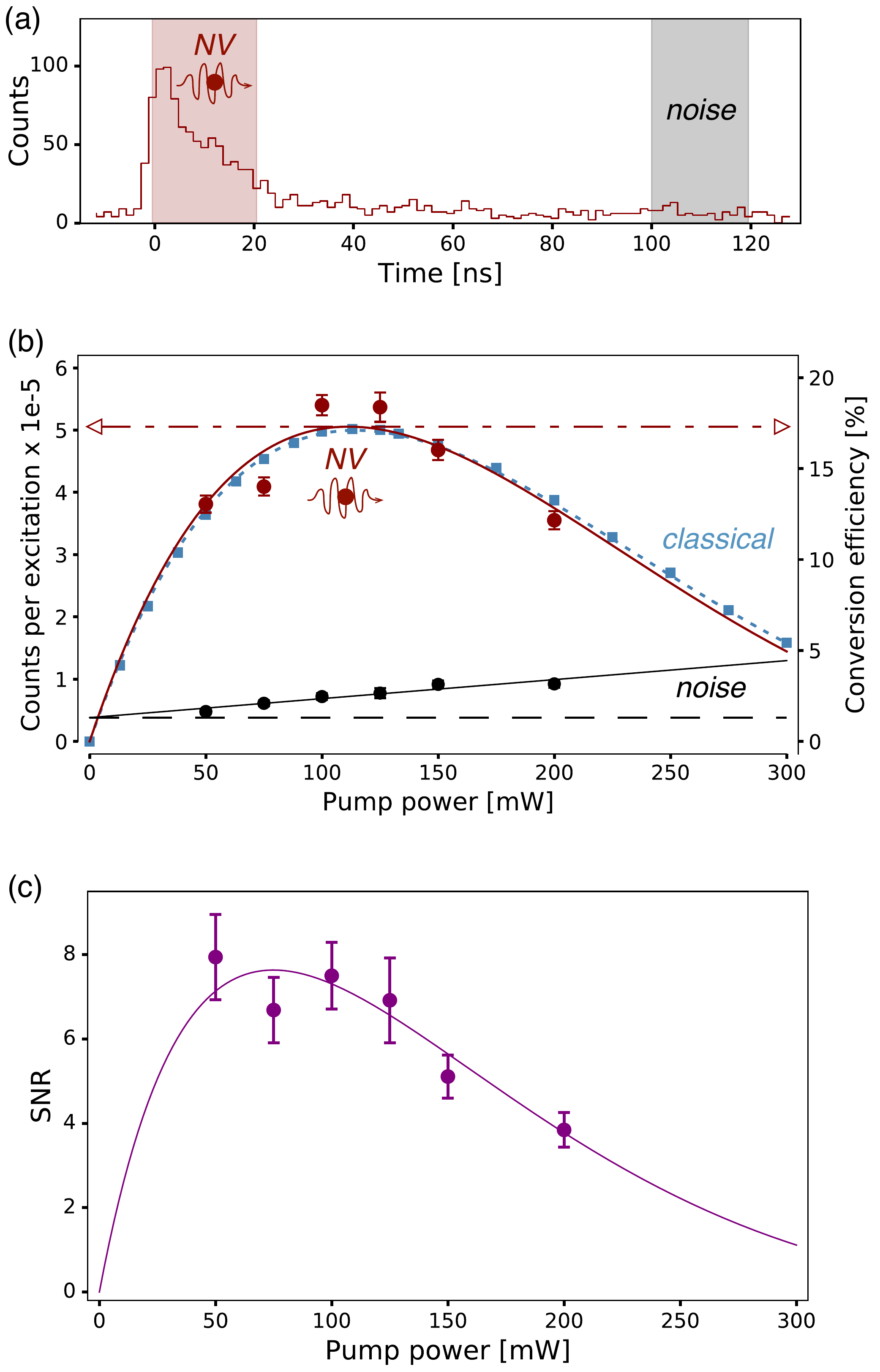}
\caption{QFC performances.
(a) Count histogram showing the time windows chosen to integrate the NV photons (red shaded area) and the noise contribution (grey shaded area).
(b) Evolution with input pump power of the probability per optical excitation to detect a telecom count coming from the NV center (dark red dots) and a noise count (black dots). 
The right axis shows the equivalence in terms of conversion efficiency. 
The evolution of the DFG measured classically is represented by the small blue dots. 
The solid and dot lines display results from fits (see main text).
The horizontal dash line represents the constant noise count level induced by detector dark counts. 
(c) Signal-to-noise ratio versus pump power. 
The solid line represents the theoretical curve obtained with parameters extracted from the fits of (b).  
The error bars correspond to one standard deviation assuming Poisson counting statistics.}
\label{fig:eff_noise}
\end{center}
\end{figure}

We now study quantitatively the quantum frequency conversion of the NV center ZPL single photons. 
To estimate the conversion efficiency, it is important to distinguish the detector events originating from the NV center down-converted photons, from the ones due to noise. 
As displayed on Figure \ref{fig:eff_noise}(a), we use temporal filtering to select the NV center counts  by integrating over a 20-ns time window following the resonant laser $\pi$-pulse exciting the defect. 
The noise reference is calculated by integrating on the same duration, but 100~ns later to avoid the collection of any photon originating from the NV center. 
Our figure of merit, from which we can estimate an expected entanglement rate, is the probability per optical $\pi$-pulse excitation, $p_c$, of recording the detection of an NV center ZPL photon.
It is calculated directly by dividing the NV center integrated counts, corrected for noise, by the number of repetitions of the experimental sequence. 
Before conversion, this probability takes the value  : $p_{c,red} = (5.7\pm 0.1)\cdot 10^{-4}$ counts/excitation, corresponding to an average photon number per pulse of $(7.1\pm 0.1)\cdot 10^{-4}$. 
The evolution of the detection probability  $p_{c,tel}$ after conversion versus the input pump power is shown in Figure \ref{fig:eff_noise}(b).
The input pump power was measured in front of the incoupling lens of the PPLN crystal, in order to take into account the imperfect effective coupling of the pump beam into the waveguide ($\approx62\%$), mainly limited by the non-optimized transmission of the incoupling lens at 1064 nm ($\approx74\%$).
The probability $p_c$ reaches a maximum of $p_{c,tel}  = (5.4 \pm 0.1) \cdot10^{-5}$ telecom photons per optical excitation, for an input pump power of $\approx110$~mW. 
The probability of detecting a telecom photon decreases for higher pump power due to the increasing contribution of the inverse process of conversion of the telecom photons to the original wavelength of 637~nm \cite{roussev_periodically_2004}.
The curve is fitted by the formula $p_{c}^{\mathrm{M}} \sin^2(\sqrt{K P_{p}} L)$ \cite{roussev_periodically_2004}, where $p_{c}^{\mathrm{M}} $ is the maximum probability of detection of a telecom photon ($p_{c}^{\mathrm{M}} = (5.1 \pm 0.2) \cdot10^{-5}$), $K$ is a constant taking into account the parameters of the conversion ($K = (9.61\pm 0.03)\cdot 10^3$ W$^{-1}$m$^{-2}$), $P_{p}$ is the input pump power and L is the length of the crystal.
The efficiency of the conversion process at the single photon level is calculated by dividing this probability by the probability of detecting a photon before conversion ($p_{c,red} = (5.7\pm 0.1)\cdot 10^{-4}$ counts/excitation), multiplied by the ratio of the QEs of the respective detectors whose values are estimated within a few percent confidence interval. 
The corresponding values can be read on the right axis of Figure \ref{fig:eff_noise}(b).
We find that the total conversion efficiency reaches $\eta_{c} \approx17\%$, corresponding to an internal conversion efficiency inside the crystal of $\approx 65 \%$. 
We have some evidence that this internal efficiency is mainly limited by imperfections in the periodicity of the poling domains \cite{SM}.
The total conversion efficiency is mainly limited by the low transmission of the narrow-band FBG filter (41\%), the imperfect fiber coupling ($\approx85\%$) and waveguide insertion and/or transmission losses ($\approx90\%$).

As a consistency check, we compare the single-photon DFG data with classical DFG data obtained by replacing the NV center photons by a 637nm continuous-wave laser and measuring the power output of the FBG on a powermeter (Thorlabs S122C). The resulting data, also plotted in Figure \ref{fig:eff_noise}(b), is observed to be in excellent agreement with the single-photon DFG data.

In QFC, the amount of noise generated plays a critical role in the success of the QFC experiment.
This is all the more true in our experiment given that the input photon rate is very low. 
As shown in Figure~\ref{fig:eff_noise}(b), the noise counts increase linearly with the input pump power. 
These counts are a combination of the constant contribution of the detector dark counts ($(0.38~\pm0.06)\cdot 10^{-5}$ counts/excitation, corresponding to a continuous rate of $190 \pm 30$ counts/s) and the pump-induced noise that is observed to be proportional to the input pump power ($(0.31~\pm~0.05)\cdot 10^{-5}$ counts/excitation per 100mW).
As mentioned before, the noise counts produced by the pump laser field are thought to arise from either Raman scattering, or SPDC. 
As demonstrated by Pelc \textit{et al.} \cite{pelc_influence_2010}, the latter noise source is enhanced by fabrication errors in the QPM grating that produces white noise spanning the transparency window of the crystal.
Both noise contributions were observed to be proportional to the input pump power \cite{zaske_efficient_2011,pelc_influence_2010}.
However, we observe for some waveguides slight deviations from this linear trend that are currently not understood 
\cite{ [ {See Supplemental Material for additional information on PPLN waveguide characteristics and details of the normalization of the autocorrelation function $g^{(2)}(\tau)$}]SM}.
Nevertheless, given that the pump and DFG wavelengths are well separated and that poling period errors are likely to be present in our waveguide given the alteration of the observed phase-matching curve \cite{SM} compared to the theoretical one  \cite{fejer_quasi-phase-matched_1992}, we are inclined to support that SPDC is the main noise contribution in our quantum frequency conversion experiment. Despite the significant remaining noise, we are able to achieve a signal-to-noise ratio (SNR) above 7 (Figure \ref{fig:eff_noise}(c)) for the direct conversion of NV center ZPL photons, thanks to the use of resonant excitation and time-resolved detection as well as efficient coupling and filtering. 
Note that a higher SNR will result when a narrower time filtering window is used (see Fig. \ref{fig:eff_noise}(a)), but at a cost of lower overall down-converted NV photon rate.
For instance, the SNR exceeds 10 if we choose a time window of 6~ns for a pump power of 75~mW.

Further increases of the SNR could come from improvements in the setup such as better non-linear crystal waveguides and optimized spectral filtering, for instance by using a FBG filter with better transmission and narrower linewidth (the FBG used here has a linewidth that is 10 times larger than that of the NV photons). Also, detectors with much lower dark count rates and better quantum efficiencies could be employed. Moreover, improvements in the collected ZPL photon rate by increased collection efficiency \cite{wan_efficient_2017} or by Purcell enhancement \cite{faraon_coupling_2012,johnson_tunable_2015,riedel_deterministic_2017,bogdanovic_design_2017} will directly enhance the SNR by the same factor. Given all these potential improvements a SNR exceeding 100 appears feasible in the near term.

\section{Photon statistics after frequency conversion}

\begin{figure}[htbp]
\begin{centering}
\includegraphics[width=8cm]{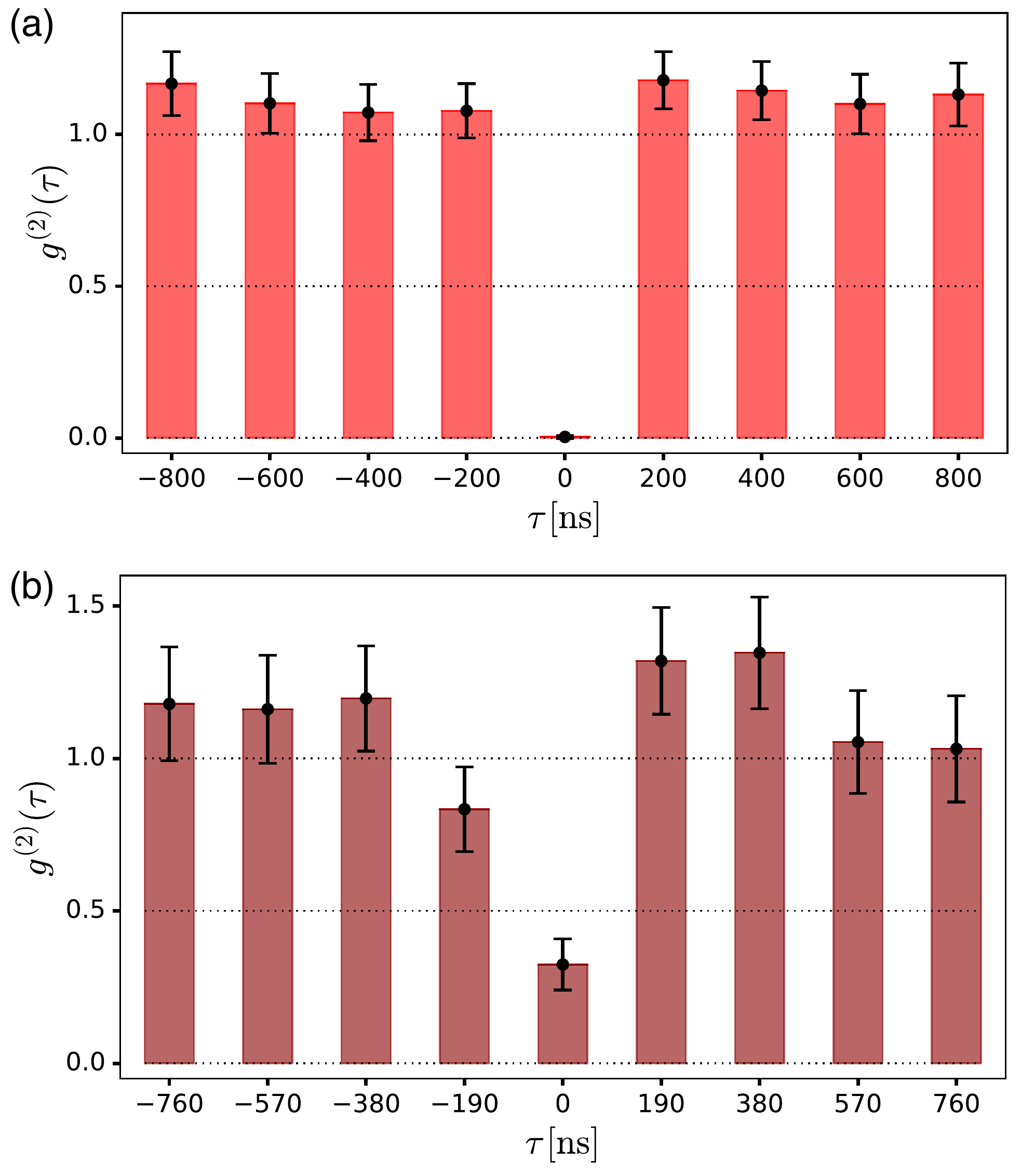}
\caption{Results of the Hanbury-Brown and Twiss pulsed experiment with the NV center ZPL photons (a) before and (b) after frequency conversion. Details of the normalization of the autocorrelation function $g^{(2)}(\tau)$ of the counts can be found in \cite{SM}.
The error bars correspond to one standard deviation. }
\label{fig:HBT}
\end{centering}
\end{figure}

Finally, we perform a Hanbury-Brown and Twiss experiment before and after conversion to investigate the effect of the down-conversion on the photon statistics. 
A PM fiber beam-splitter is added after the narrow FBG and its two outputs are directed to single-photon detectors at the appropriate wavelengths, each one connected to a channel of the time-tagged device. 
In order to limit the occurrence of spectral diffusion, spin flips and ionization of the NV center, the experiment is divided into sequences of 15 resonant optical excitation $\pi$-pulses. In between these sequences, the NV spin is re-initialized by optical pumping. Furthermore, after every 250 repetitions of this sequence a charge state and resonance check is performed~\cite{bernien_heralded_2013}.
The input pump power was set to 100~mW. 
From the count histograms associated with both detector channels, we construct the temporal coincidence count histogram. 
In order to get more coincidence detections in our statistics, we increase the temporal window to 25~ns of integration after the beginning of the pulse.
The resulting $g^{(2)}(\tau)$ function centered on zero delay is displayed on Figure \ref{fig:HBT} for photons before and after down-conversion. For both cases, clear anti-bunching behaviour is observed~\cite{beveratos_room_2002}. 
The bars are separated by the delay between the laser excitation pulses, being of 200~ns and 190~ns for the 637~nm and the telecom photons, respectively.
Since the coincidence rate evolves as the square of the input rate, the coincidence counts are roughly 13 times lower after conversion than before despite a much longer integration time ($\approx$43h compare to 2h for the data recorded at 637 nm). 
The values $g^{(2)}(\tau) >1 $ for the side peaks are well-known signatures of photon bunching associated with an effective blinking effect due to shelving in a long-lived metastable state \cite{beveratos_room_2002}.

The $g^{(2)}(0)$ before conversion reaches $0.003 \pm 0.002$, showing the excellent purity of the NV center as a resonant single photon source (SNR $\approx530$).
The non-zero value is well explained by the dark counts of the APD ($\approx 10$ dark counts/s).  After conversion, the $g^{(2)}(0)$ increases to $0.32 \pm 0.08$ due to the noise produced during the QFC process, while still remaining under the threshold of 0.5 expected for single emitter. 
Importantly, this value for $g^{(2)}(0)$ is consistent with the expected $g^{(2)}(0) \approx 0.35$ one would get with a SNR of $\approx4.2$ extracted from the count histograms (see \ref{sec:eff_noise} and \cite{SM}) \cite{beveratos_room_2002}. 
(Note that the SNR in this experiment is lower because the photon count rate is divided by two due to the presence of the beam splitter, but detector dark counts remain constant.) 
We conclude from these measurements that, although degraded by the pump noise, the single-photon statistics of the NV center persist after the down-conversion process. 

\section{Conclusion and outlook}

Our experiment demonstrates the down-conversion of single  NV center spin-selective photons at 637~nm to 1588~nm in the telecom L-band. 
Our direct down-conversion reaches a total efficiency of 17\% at a signal-to-noise ratio of $\approx$7, limited by detector dark counts and pump-induced noise that we attribute to unwanted SPDC processes taking place in the nonlinear crystal. We were able to observe clear photon anti-bunching without background noise subtraction in the autocorrelation function of the light after frequency conversion.

Considering the rate of entanglement generation between distant NV centers in a quantum network, frequency conversion with the current setup configuration would already be advantageous at a separation corresponding to 2.6km of fiber for a two-photon-based entangling protocol as used in Refs~\cite{barrett_efficient_2005,bernien_heralded_2013,hensen_loophole-free_2015}. 
An additional advantage of down-conversion is that same-frequency telecom photons - a key requirement for entangling protocols - can be generated out of NV centers that emit at different frequencies due to strain, thus removing the need for dc Stark tuning~\cite{bernien_heralded_2013}. The results presented here are therefore a key step towards future implementations of quantum communication networks based on NV centers in diamond.

\section*{Acknowledgements}
The authors are grateful to Florian Kaiser, Peter Humpreys, Florent Doutre, Christoph Becher, and S\'ebastien Tanzilli for fruitful discussions. 
We thank Norbert Kalb for his careful reading of the manuscript. 
We thank Siebe Visser, Jelle Haanstra, Raymond Schouten, Marijn Tiggelman and Raymond Vermeulen for technical support.  We thank Ronan Gourges and Single Quantum company staff for technical discussions and support. We acknowledge support from the Netherlands Organisation for Scientific Research (NWO) through a Vici grant and the European Research Council through a Synergy Grant.

\bibliography{ref_QTelecom_final.bib}

\end{document}